\documentclass[pra,aps,amssymb,showpacs]{revtex4}
\usepackage{graphicx}
\newcommand{\ket}[1]{|#1\rangle}
\newcommand{\bra}[1]{\langle#1|}

\newcommand{\ketbra}[2]{|#1\rangle\langle #2|}

\begin{document}

\title{Entanglement purification without controlled-NOT gates by using the natural dynamics of spin chains}
\author{Koji Maruyama$^{1,2}$ and Franco Nori$^{1,2,3}$}
 \affiliation{$^1$Advanced Science Institute, The Institute of Physical and Chemical
Research (RIKEN), Wako-shi 351-0198, Japan\\
$^2$ CREST, Japan Science and Technology Agency, Kawaguchi, Saitama 332–0012, Japan\\
$^3$ Center for Theoretical Physics, Physics Department, Center for the Study of Complex Systems,
University of Michigan, Ann Arbor, MI 48109-1040, USA}
\begin{abstract}
We present a simple protocol to purify bipartite entanglement in spin-1/2 particles by utilizing
only natural spin-spin interactions, i.e. those that can commonly be realized in realistic
physical systems, and $S_z$-measurements on single spins. Even the standard isotropic Heisenberg
interaction is shown to be sufficient to purify mixed state entanglement. Our protocol does not
need controlled-NOT (CNOT) gates that are very hard to implement experimentally. This approach
could be useful for quantum information processing in solid-state-based systems.
\end{abstract}
\date{\today}
\pacs{03.67.Bg, 03.67.Pp, 75.10.Pq}
\maketitle

\section{Introduction}
Entanglement purification is one of the most important tasks in quantum information processing
\cite{bennett96a,bennett96b,deutsch96,dur07}. It is a process to extract strongly entangled pairs
out of initially weakly entangled ones using local operations and classical communication. By
repeating the purification process, (near-) maximal entanglement can be obtained. Such a task is
indispensable because maximally entangled states are an irreplaceable resource for many important
protocols, such as quantum cryptography, quantum teleportation, and quantum repeaters.
Unfortunately, as entanglement can only be generated by direct physical interactions, the coherent
physical transfer of quantum states is necessary. However, such a transport is always fraught with
difficulties due to interactions with the environment, which lowers the quality of entanglement.
Therefore, a number of quantum information processing tasks rely on the feasibility of
entanglement purification, which is the key process to distill arbitrarily high entanglement out
of degraded states.

Most well-known entanglement purification protocols for bipartite two dimensional systems were
proposed by Bennett et al. \cite{bennett96a,bennett96b} (the \textit{BBPSSW} protocol) and Deutsch
et al. \cite{deutsch96} (the \textit{DEJMPS} protocol). Both operate on two pairs shared between
two separated parties, Alice and Bob, to enhance entanglement in one of the pairs. An operation
commonly used in these protocols is a controlled-NOT (CNOT) gate applied at both ends of the
channel. A CNOT flips one of the spins (target spin) if the other (control spin) is in $\ket{1}$,
and leaves the target spin unchanged if the control spin is in $\ket{0}$, where $\ket{0}$ and
$\ket{1}$ are some orthogonal vectors. Yet, in general, carrying out a CNOT operation is one of
the toughest challenges in experiments. This is why there have been only a few experiments of
purification, mainly with photons \cite{pur_experiments} and one with trapped ions
\cite{reichle06}. Throughout this paper we will consider spin-$1/2$ particles as a representative
of two-level systems because our main interest is in the behavior of entanglement in spin chains
with standard spin-spin interactions.

The main aim of this paper is to show that entanglement purification is possible only with
unmodulated natural interactions between spins and simple $S_z$-measurements on single spins, as
long as the fidelity of the initial state to a Bell state is larger than $1/2$. A physical
intuition behind this idea is that measuring a subset of spins would project the remaining pairs
onto a more strongly entangled state since there are constructive and destructive interference of
magnons in a spin chain during its dynamical evolution \cite{sougato03, maruyama07, burgath07}.

Here, there is \textit{no} need of mapping quantum information to different physical systems, e.g.
photons. Besides, there is \textit{no} need of artificial multi-qubit gate operations, such as
CNOT. With an isotropic Heisenberg interaction, the implementation of a CNOT has to involve at
least using two two-spin operations \cite{loss98}, that is, the accumulation of errors is very
likely to occur. From the pragmatic point of view, we naturally wish to minimize the number of
artificial controls. Our proposal simplifies the necessary manipulations, compared with those in
experiments \cite{reichle06}, and also theoretical proposals for solid state circuits, such as
\cite{taylor05}. It does not entail even any single spin operations after the first round. Several
analyses have been done, for example, for quantum state transfer with a chain of spin-1/2
particles \cite{sougato03} or harmonic oscillators \cite{martin04}, elementary gate operations
\cite{yung04} and cloning transformation \cite{dechiara05} under similar motivations, however,
ours performs an invaluable, and practically very important, quantum information processing task
with minimal control.

\section{Entanglement purification with spin chains}
\begin{figure}
\begin{center}
\includegraphics[scale=1]{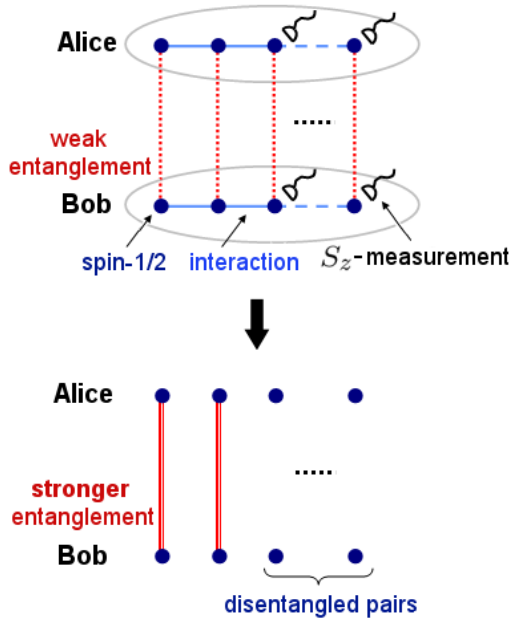}
\caption{(Color online) Entanglement purification using spin chains. Alice and Bob initially share
$n$ pairs of weakly entangled spins forming two spin chains. They let the spin chains evolve under
a standard spin-spin interaction and perform single-spin measurements on a subset of the spins.
Entanglement purification succeeds if the remaining $m$ pairs contain higher entanglement than the
initial state.}\label{purification}
\end{center}
\end{figure}

A general scenario of entanglement purification by spin chain dynamics is depicted in Fig.
\ref{purification}. Alice and Bob share $n$ pairs of weakly entangled spins forming two spin
chains so that Alice's $j$-th spin is entangled with Bob's $j$-th spin for all $j$. Each chain
will evolve due to the exchange interaction between nearest-neighboring spins. We will primarily
consider the isotropic Heisenberg interaction, however, let us assume the Hamiltonian in a generic
anisotropic form as
\begin{equation}\label{anisotropichamiltonian}
H=-\sum_{\langle i,j\rangle} (J_x S_i^x S_j^x + J_y S_i^y S_j^y + J_z S_i^z S_j^z),
\end{equation}
where $J_k$ and $S_i^k\,(k=x,y,z)$ are the coupling strengths and the standard $SU(2)$ spin
operators for the $i$-th spin, respectively. After a certain time lapse they measure $n-m$ spins
each in the $\{\ket{\!\uparrow},\ket{\!\downarrow}\}$ basis, turning off the interaction, to
enhance or reduce entanglement in the remaining $m$ pairs compared with that of the initial state.
All active controls we consider here are one switching-on and one switching-off of the exchange
interaction Eq. (\ref{anisotropichamiltonian}), and single-spin measurements in the
$\{\ket{\!\uparrow},\ket{\!\downarrow}\}$ basis. Single spin operations can be added to the list
of available controls, however, the fewer controls, the better. In order to avoid complications
due to residual correlations between unmeasured pairs, we consider cases in which only one pair is
left unmeasured after a run of the protocol. We will hereafter denote vectors $\ket{\!\uparrow}$
and $\ket{\!\downarrow}$ by $\ket{1}$ and $\ket{0}$, respectively, and take $\hbar=1$.

\begin{figure}
\begin{center}
\includegraphics[scale=.5]{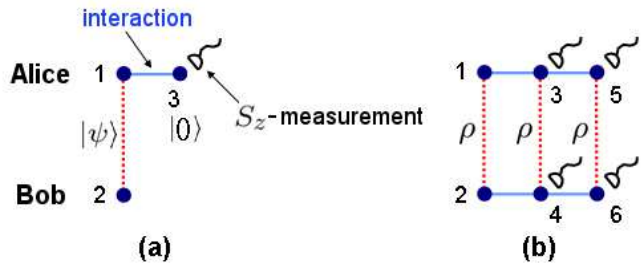}
\caption{(Color online) Configurations to obtain one pair with higher entanglement by spin chain
dynamics. (a) If the initial state of a pair is pure, an ancillary spin suffices. (b) If the pairs
are initially in a mixed state, we need three pairs, two of which are to be measured. The blue
lines and red dashed lines indicate interaction and entanglement, respectively, and the
measurement is in the $\{\ket{0},\ket{1}\}$ basis.}\label{config}
\end{center}
\end{figure}

If the initial state of the pair is a partially entangled pure state \cite{note1}, then a
maximally entangled pair can be attained by simply attaching only one spin to the pair locally and
letting the two spins evolve under the isotropic Heisenberg interaction, that is, Eq.
(\ref{anisotropichamiltonian}) with $J_x=J_y=J_z=J$. Suppose that the initial state of the pair is
given by $\ket{\psi}=\alpha\ket{00}+\beta\ket{11}$, where $\alpha, \beta \in \mathbb{R}$,
$\alpha^2+\beta^2=1$ and $\alpha<\beta$ (the case of $\alpha>\beta$ can be treated in the same
fashion). See Fig.~\ref{config}(a) for a schematic diagram and the numbering of the spins. The
three-spin state at time $t$ can be computed as
\begin{eqnarray}\label{pure_time_evolution}
\ket{\psi(t)} &=& \alpha e^{iJt/4}\ket{000} \nonumber \\
& & + \beta e^{-iJt/4}\left(i\sin \frac{Jt}{2} \ket{011} + \cos\frac{Jt}{2}\ket{110}\right),
\end{eqnarray}
where spins are ordered according to the numbering in Fig.~\ref{config}. If an $S_z$-measurement
on the spin `3' at $Jt_\mathrm{max}=2\cos^{-1}(\alpha/\beta)$ gives the outcome 0, the state of
the first pair (spins `1' and `2') will be projected onto a maximally entangled state with
probability $2\alpha^2$. The net result stays the same even with the (isotropic) antiferromagnetic
Hamiltonian, i.e. $J<0$.

This process can be seen as equivalent to local filtering \cite{gisin96}, in which a spin and an
ancilla evolve under some unitary operation, and then a measurement is performed on the ancilla.
For specific measurement outcomes the pair becomes closer to the maximally entangled state.

Let us move on to the case in which the initial state of the pairs is mixed. In order to simplify
the discussion, we shall assume that each pair is in a Werner state \cite{werner89}, that is
\begin{equation}\label{werner_state}
\rho_W=F\ket{\Phi^+}\bra{\Phi^+}+\frac{1-F}{3}(\ket{\Phi^-}\bra{\Phi^-}+\ket{\Psi^+}\bra{\Psi^+}+\ket{\Psi^-}\bra{\Psi^-}),
\end{equation}
where $\ket{\Phi^\pm}=(\ket{00}\pm\ket{11})/\sqrt{2}$ and
$\ket{\Psi^\pm}=(\ket{01}\pm\ket{10})/\sqrt{2}$ are the maximally entangled Bell states, and $F$
is the fidelity between $\rho_W$ and $\ket{\Phi^+}$. The reasoning for this assumption is that any
mixed state can be transformed into a Werner state with random bilateral local unitary operations
(twirling) without lowering its fidelity with a Bell state, thus showing the feasibility of
entanglement purification for Werner states is sufficient in principle \cite{bennett96a}. In what
follows, we shall use the BBPSSW protocol as a reference for comparisons because it is easier than
others to analyze.

It turns out that higher entanglement can never be attained by starting with two pairs of spins,
$\rho_W^{\otimes 2}$, regardless of the combinations of values of $(J_x, J_y, J_z)$ in the
Hamiltonian. Nevertheless, entanglement in a pair can be enhanced if there are two extra pairs,
i.e. three pairs in total as in Fig.~\ref{config}(b). The reason why this approach fails with two
pairs will be explained later.

\begin{figure}
\begin{center}
\includegraphics[scale=.6]{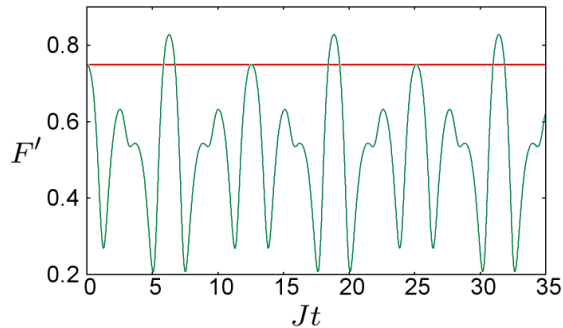}
\caption{(Color online) The fidelity as a function of the measurement time, with the initial
fidelity equal to 0.75, indicated by the red (horizontal) line.}\label{fidelity_change}
\end{center}
\end{figure}

\begin{figure}
\begin{center}
\includegraphics[scale=.5]{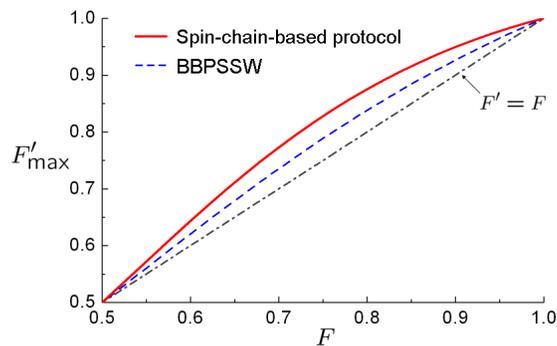}
\caption{(Color online) Maximum attainable fidelity $F_\mathrm{max}^\prime$ between the resulting
state and a Bell state. The top (red) solid line represents the maximum fidelity after a run of
our protocol with three pairs. The dashed (blue) line shows the fidelity change by the BBPSSW
protocol. The bottom (black) dash-dotted line represents $F^\prime =F$.}\label{fidelityplot}
\end{center}
\end{figure}

If there are three pairs, the resulting fidelity does exceed the initial one. An example of the
`time evolution' of the fidelity is plotted in Fig.~\ref{fidelity_change}. In this plot, the
isotropic nearest-neighbor interaction, $H=J\sum \vec{S}_i\cdot\vec{S}_j$, is assumed and the time
$t$ indicates the measurement time. With the initial state $\rho_W$ above, the measurement
outcomes $(0,0,1,1)$ or $(1,1,0,0)$ at sites $\langle 3,4,5,6\rangle$ or $\langle 1,2,5,6\rangle$
lead to a successful fidelity increase (Hereafter angular brackets $\langle\cdot\rangle$ denote
the spin sites). The maximum fidelity attainable can be calculated analytically as
\begin{equation}\label{max_fidelity}
F^\prime_\mathrm{max}=\frac{16-53F+118F^2}{59-106F+128F^2},
\end{equation}
which is always larger than $F$ if $F>1/2$ and is achieved at times $Jt=(2n+1)2\pi \,
(n=0,1,2,...)$. Figure \ref{fidelityplot} shows the comparison between the maximum reachable
fidelity by our protocol and that by the BBPSSW. The fidelity increase in the spin-chain-based
protocol is approximately twice as large as that by BBPSSW, which is quite significant despite the
simplicity of the protocol. There are other combinations of outcomes, $(0,0,0,0)$ and $(1,1,1,1)$,
which give a fidelity increase, but we exclude this possibility because the increase is small
(roughly half of the BBPSSW) while the probability is slightly higher than the above case.

What about the necessary time precision for the protocol? The duration (FHWM) $\Delta t$ to have a
fidelity increase larger than $(F^\prime_\mathrm{max}-F)/2$ depends on $F$, but it can be computed
to be $J\Delta t<0.5$ for a wide range of $F$, i.e. $0.61<F<0.94$. With $J\sim 1\mu$eV in, e.g.
\cite{petta05}, this condition gives $\Delta t < 300$ ps, which is much longer than the
experimentally feasible accuracy of $\sim10$ ps.

Another nice feature of this protocol is that the purified pair stays in the Werner state with the
new fidelity $F^\prime_\mathrm{max}$, i.e.
$\rho^\prime=F^\prime_\mathrm{max}\ket{\Phi^+}\bra{\Phi^+}+(1-F^\prime_\mathrm{max})/3(\ket{\Phi^-}\bra{\Phi^-}+\ket{\Psi^+}\bra{\Psi^+}+\ket{\Psi^-}\bra{\Psi^-})$.
This means that further twirling operations are not necessary when iterating the protocol with
three new pairs in $\rho^\prime$, unlike the BBPSSW.

\section{Discussion}
Let us now discuss the case where there are only two pairs of spins for our scheme. As we have
mentioned above, two pairs of a Werner state cannot be purified if there are no extra single spin
operations. We sketch its reasoning and show that the protocol can be as efficient as the BBPSSW
when single spin operations are available in addition to a single switching on and off of the
two-spin interaction of the form of Eq. (\ref{anisotropichamiltonian}).

The Hamiltonian Eq.~(\ref{anisotropichamiltonian}) has its invariant subspaces spanned by the
following combinations of Bell-product states, namely, $(\Phi^+\Phi^-,\Phi^-\Phi^+)$,
$(\Phi^+\Psi^\pm,\Psi^\pm\Phi^+)$, $(\Psi^+\Psi^-,\Psi^-\Psi^+)$,
$(\Psi^\pm\Phi^-,\Phi^-\Psi^\pm)$, and $(\Phi^+\Phi^+,\Phi^-\Phi^-,\Psi^+\Psi^+,\Psi^-\Psi^-)$.
Note that these subspaces are independent of the values of the coupling strengths $J_k$
$(k=x,y,z)$, because all terms in Eq.~(\ref{anisotropichamiltonian}) commute with each other, thus
simultaneously diagonizable regardless of $J_k$'s. In the following, the dominant component in the
initial Werner state is assumed to be $\Phi^+$ and we will look at the resulting state of one of
the pairs when the outcomes of the $S_z$-measurement on the other pair are the same. We do not
lose any generality by such assumptions thanks to the symmetry of the Hamiltonian and the Werner
state.

The initial state $\rho_W^{\otimes 2}$ is diagonal in the Bell-product basis and there are three
values of the weight for those Bell-product components. Let us write those weights as $w_L=F^2,
w_M=F(1-F)/3$, and $w_S=(1-F)^2/9$. Since the components in the same rank-2 subspace above have an
equal weight, $\rho_W^{\otimes 2}$ is equivalent to an identity operator in each rank-2 subspace.
Thus the time evolution of entanglement is determined by that of components in the rank-4
subspace. If we evaluate the fidelity with respect to $\Phi^+$, its maximum is obtained when the
weight of $\Phi^+\Phi^+$ comes back to $w_L$. The (unnormalized) state of spins $\langle
1,2\rangle$ after measuring (0,0) or (1,1) at sites $\langle 3,4\rangle$ then becomes
$w_L\ketbra{\Phi^+}{\Phi^+}+w_S\ketbra{\Phi^-}{\Phi^-}+
w_M(\ketbra{\Phi^+}{\Phi^+}+\ketbra{\Phi^-}{\Phi^-}+\ketbra{\Psi^+}{\Psi^+}+\ketbra{\Psi^-}{\Psi^-})
+w_S(\ketbra{\Psi^+}{\Psi^+}+\ketbra{\Psi^-}{\Psi^-})$. The fidelity between this state and
$\Phi^+$ is $(w_L+w_M)/(w_L+4w_M+3w_S)=F$, hence there is no fidelity increase. The same logic can
be applied straightforwardly to cases with different expected outcomes and a different reference
Bell state for fidelity evaluation.

If we allow anisotropy for the interaction, entanglement purification can be achieved with as big
fidelity increase as in the BBPSSW protocol with the help of extra single spin operations. This is
simply because a CNOT can be decomposed into a sequence of single spin operations and one-time
activation of the XY or Ising-type interaction \cite{schuch03}. But with the isotropic Heisenberg
interaction, such a decomposition is known to be impossible \cite{burkard99}.

The effect of the Heisenberg Hamiltonian is quite different when there are three pairs and its
invariant subspaces can no longer be described neatly by the Bell states. Physically, this
difference can be attributed to the behavior of spinless fermions that appear after the
Jordan-Wigner transformation. When there are only two spins in a chain the fermion propagates (or
hops) freely between sites, hence the simple dynamics described above. On the contrary, in a
three-spin chain, the fermions interact with each other and there can be some nontrivial
interferences between different modes that result in the concentration of entanglement.

Despite such an intuitive picture, the application of the Jordan-Wigner transformation to our
problem makes the whole calculation quickly untractable. Instead, here we attempt to characterize
the three-pair Hamiltonian in terms of the ability of making correlations between Bell pairs. Such
an analysis is useful because entanglement purification can be seen as a process to extract
information on a given pair by correlating other pairs to it. The BBPSSW protocol makes use of the
correlating power of the bilateral CNOT, which correlates two Bell states as
\begin{equation}\label{cnot_correlation}
\ket{\Psi_{kl}}\ket{\Psi_{mn}}\longrightarrow \ket{\Psi_{k,l+n}}\ket{\Psi_{k+m,n}},
\end{equation}
where $\Psi_{00}, \Psi_{01}, \Psi_{10}, \Psi_{11}$ are in our notation $\Phi^+, \Phi^-, \Psi^+,
\Psi^-$, respectively, and additions in the indices are in modulo 2. If a bilateral CNOT is
applied on a randomly chosen Bell pair and another pair in a known Bell state, the mutual
information between these two pairs is increased from 0 bits to 1 bit. Similarly, the time
evolution operator $U(Jt=2\pi)=\exp[-2\pi i\sum \vec{S}_i\cdot\vec{S}_j]$ for three pairs
increases the mutual information (in the Bell basis), between a given unknown pair and the other
two, from 0 to 1.48 bits, while $U=\exp(-iHt)$ generates no correlation between pairs in the
two-pair case. The correlation generated between pairs then gives (partial) information on the
unmeasured pair, enabling us to single out pairs containing higher entanglement.

A price we have to pay for the simplicity and the larger fidelity increase is the probability of
success, which is about 2/5 of that of the BBPSSW. Let us make a rough estimation of the number
$l$ of initial pairs to obtain a final pair achieving a desired fidelity increase for BBPSSW
($l_{B}$) and for our spin-chain-based protocol ($l_\mathrm{SC}$). These can be written as
\begin{equation}
l_{B}=\left(\frac{2}{p_B}\right)^{r_B}\,\,\, \mathrm{and}\,\,\,
l_\mathrm{SC}=\left(\frac{3}{p_\mathrm{SC}}\right)^{r_\mathrm{SC}},
\end{equation}
where $p_B$ ($p_\mathrm{SC}$) and $r_B$ ($r_\mathrm{SC}$) are the success probability per run and
the number of rounds (`rounds' as in a sport tournament), respectively. We approximate
$p_\mathrm{SC}$ by its average over the range $1/2<F<1$ for simplicity as it does not vary much.
Also, by expanding the form of $F^\prime$ with respect to $F$ around 1, we see that
\begin{eqnarray}\label{r_B}
r_{B} &\simeq & \left(\log\frac{2}{3}\right)^{-1}\cdot\log\left(\frac{1-F_f}{1-F_i}\right)
\nonumber \\
&\simeq & \left(\log\frac{2}{3}\right)^{-1}\cdot\log\left(\frac{11}{27}\right)\cdot r_\mathrm{SC}
=  2.21r_\mathrm{SC}
\end{eqnarray}
for a fidelity change from $F_i$ to $F_f$. With these approximations, we find that
\begin{equation}\label{l_comparison}
\frac{l_\mathrm{SC}}{l_B}\simeq 1.16^{r_\mathrm{SC}}
\end{equation}
holds for any fidelity increase. This might make our protocol appear less attractive than the
BBPSSW. However, more important is the number of multi-spin operations the resulting pair
experiences. In our scheme this number is $r_\mathrm{SC}$, which is only 23\% of $2r_\mathrm{B}$,
the operations needed for the standard CNOT-based BBPSSW protocol. Besides, the difference in $l$
is rather small. Even if $r_\mathrm{SC}=5$, which corresponds to a fidelity increase from 0.75 to
0.990, $l_\mathrm{SC}/l_B$ is only 2.07. We therefore wish to state that the efficiency of our
protocol is still notable, considering that \textit{only} natural evolutions and an initial
one-time twirling are involved.

Before summarizing, let us briefly note the effect of anisotropy. Our protocol does not work with
the XY or Ising interaction. Yet, a small amount of anisotropy is useful to purify mixed state
entanglement. For example, suppose that there is an antisymmetric exchange interaction, also known
as the Dzyaloshinskii-Moriya interaction, in addition to the XY interaction as
\begin{equation}
H=-J\sum (S_i^x S_j^x+S_i^y S_j^y) + \vec{d}\cdot\vec{S}_i\times\vec{S}_j,
\end{equation}
where $\vec{d}$ is the coupling vector that reflects the anisotropy. If a set of three pairs, each
of which is in the Werner state with fidelity $F=0.75$, evolves under the Hamiltonian with $J=1$
and $\vec{d}=\{0.1,0,0\}$, the fidelity can be as high as 0.81 when two pairs are measured at
$t\simeq 352$.

\section{Summary and outlook}
We have shown that entanglement purification is possible with only the free dynamics of spin
chains and simple $S_z$-measurements on single spins. Arbitrarily high entanglement can be
obtained by repeating the process. The amount of operations is minimal, in the sense that single
spin operations (twirling) are needed only once for the entire process and only one activation of
the inter-spin interaction is required for each run. Hence the error accumulations due to
artificial controls should be much less likely compared with conventional schemes. More detailed
and quantitative comparisons of errors under realistic conditions would thus be one of the
important future research topics.

The proposed scheme might also be experimentally realizable in the near future as the number of
spins in this protocol is not large. Candidate physical systems would be, for example,
semiconductor quantum dots \cite{loss98,petta05,sarma05}, superconducting qubits \cite{you05},
atoms trapped in optical lattices \cite{mandel03}, etc., where the inter-qubit couplings are often
described by Heisenberg-type interactions.

If we look at our protocol from a more general perspective, it can be understood as an execution
of a quantum information processing task by a natural time evolution and simple measurements.
While we have considered only one specific task in this paper, it would be very intriguing if more
complicated tasks could be performed with simple multi-particle interactions.

\section*{Acknowledgments}
KM is grateful to Martin Plenio for a careful reading of the manuscript. Some calculations were
performed with the RSCC system of RIKEN. This work was supported in part by the National Security
Agency (NSA), the Army Research Office (ARO), the Laboratory for Physical Sciences (LPS), the
National Science Foundation (NSF) grant No. EIA-0130383, the CREST program of the Japan Science
and Technology Agency (JST), the Japan Society for the Promotion of Science Core-to-Core
(JSPS-CTC) program, and the JSPS-RFBR grant 06-02-91200.


\begin{thebibliography}{99}
\bibitem{note1}
In this case, `entanglement concentration' is a more appropriate term than `purification'.
\end{thebibliography}
\end{document}